\documentclass[a4paper]{quantumarticle}
\pdfoutput=1
\usepackage[utf8]{inputenc}
\usepackage[english]{babel}
\usepackage[T1]{fontenc}
\usepackage{caption}
\usepackage{subcaption}
\usepackage{cite}
\usepackage{amsmath}
\usepackage{hyperref}

\usepackage{tikz}
\usepackage{lipsum}

\begin{document}

\title{Simulating the Quantum Rabi Model in Superconducting Qubits at Deep Strong Coupling}
\author{Noureddine Rochdi}
\affiliation{LPHE-Modeling and Simulation, Faculty of Sciences, Mohammed V University in Rabat, Rabat, Morocco.}
\email{noureddine\_rochdi@um5.ac.ma}
\homepage{https://www.researchgate.net/profile/Noureddine-Rochdi-3}
\orcid{0000-0002-2445-2701}
\author{Atta ur Rahman}
\email{attaphy57@mails.ucas.ac.cn}
\homepage{https://www.researchgate.net/profile/Atta-Ur-Rahman}
\affiliation{School of Physical Sciences, University of Chinese Academy of Sciences, Yuquan Road 19A, Beijing 100049, China}
\orcid{0000-0002-2445-2701}

\author{Rachid Ahl Laamara}
\email{r.ahllaamara@um5r.ac.ma}
\homepage{https://www.researchgate.net/profile/Rachid-Laamara}
\orcid{0000-0003-0290-4698}
\affiliation{LPHE-Modeling and Simulation, Faculty of Sciences, Mohammed V University in Rabat, Rabat, Morocco.}

\author{Mohamed Bennai}
\email{mdbennai@yahoo.fr}
\affiliation{Quantum Physics and Magnetism Team, LPMC, Faculty of Sciences BenM’sick, Hassan II University of Casablanca, Morocco}
\homepage{https://www.researchgate.net/profile/Mohamed-Bennai}
\orcid{0000-0003-1985-4623}
\maketitle

\begin{abstract}
The Quantum Rabi model serves as a pivotal theoretical framework for elucidating the nuanced interplay between light and matter.  Utilizing circuit quantum electrodynamics on a chip, we address the challenge of achieving deep strong coupling in Quantum Cavity Electrodynamics (cQED). Despite progress in superconducting circuits and trapped ions, experimental realization has been limited to spectroscopy. Our focus is on a transformative digital quantum simulation, employing Trotterization with an augmented number of steps to deconstruct a complex unitary Hamiltonian. This approach showcases the benefits of digital techniques within superconducting circuits, offering universality, flexibility, scalability, and high fidelity. Our goal is to demonstrate deep strong coupling in cQED and understand the advantages of digital methods, particularly in coherent measurement during time evolution with varying photon counts in resonators. This opens avenues to leverage quantum mechanics for overcoming hardware limitations.
\end{abstract}
\\ 
\section{Introduction}
In the realm of ultra-strong coupling \cite{example} and deep strong coupling \cite{exampl} in quantum technologies\cite{cv}, researchers are delving into uncharted territories that challenge our fundamental understanding of quantum systems. Ultra-strong coupling \cite{merci} refers to a regime, where $g^R$ (the coupling intensity), reaches to a comparable or even exceeds the energy associated with bosonic mode, represented by $\omega_b^R$. This novel domain opens up avenues for exploring exotic quantum phenomena that were previously inaccessible \cite{cnrt}. The dynamics in ultra-strong coupling regimes may lead to unconventional behaviors, offering unique opportunities for manipulating quantum states and developing applications in quantum information processing. Besides this, the exploration of deep strong coupling unveils another layer of complexity in quantum systems \cite{a,b,c}. In this regime, the interaction between a qubit and a quantized harmonic \cite{mac,macs}mode becomes significantly pronounced, leading to intricate and rich dynamics. The interplay between the Jaynes-Cummings configurations with quantum Rabi model, in particular, about the rotating wave approximation with weak coupling, takes on new dimensions when subjected to deep strong coupling \cite{digital}. Researchers are actively investigating the implications of these phenomena on quantum protocols\cite{example,exampl,merci,a,b,c,digital,biblatexsubmittingtothearxiv,arxivpdfoutput,roch}, seeking to harness the potential of deep strong coupling for the development of advanced quantum technologies. The challenges and opportunities presented in the domain of deep  and ultra-strong coupling are pivotal in shaping the future landscape of quantum science and technology.

Several decades ago, physicists, most notably Richard Feynman \cite{examplecitation}, recognized the formidable challenge of simulating quantum mechanics. A critical bottleneck emerged in the form of the substantial processor memory needed to store the quantum state of a sizable physical system, a challenge that has since been addressed. The quantum state is characterized by parameters that grow exponentially with the system's size, typically influenced by the number of particles or degrees of freedom involved. To tackle this challenge, Feynman proposed a solution: simultaneous implementation of simulations alongside the evolution of a quantum computers, and hardware. This approach would lead to an exponential explosion, potentially providing valuable insights and positive outcomes in overcoming the computational barriers associated with quantum system simulations \cite{examples}.

In recent years, interest in the significance of quantum simulation has burgeoned for two primary reasons \cite{examplecitation}. For example, it offers a multitude of potential applications in physics, chemistry \cite{noured}, and even biology \cite{grand}, with a focus on condensed matter physics \cite{cite}. Quantum simulation provides a unique platform to model and understand complex quantum systems that are challenging to simulate using classical methods. This capability holds promise for unraveling the mysteries of materials\cite{canda,cande,candea}, chemical reactions\cite{candat}, and biological processes at the quantum level. Besides, quantum simulation plays a crucial role in exploring and harnessing quantum technologies. As we venture into the era of quantum computing, quantum simulators \cite{grantis} serve as invaluable tools for testing algorithms, verifying quantum hardware, and simulating quantum dynamics. They provide a controlled environment to study quantum phenomena, paving the way for advancements in quantum information processing, optimization \cite{granda}, and cryptography \cite{grande}. In essence, the burgeoning significance of quantum simulation lies in its potential to revolutionize our understanding of fundamental sciences and contribute to the enhancement of modern quantum technologies.

In the realm of quantum information, the traditional classification of quantum simulators distinguishes between analog and digital approaches \cite{r}. Analog quantum simulators involve devices that are designed to replicate a specific Hamiltonian, effectively emulating the quantum system under consideration. However, introducing tunability to these simulators expands their capabilities beyond a fixed Hamiltonian. Tunable analog devices provide the flexibility to address a broader class of quantum problems, offering a more versatile and adaptable approach. The inclusion of tunability is particularly significant as it allows for the exploration of diverse quantum phenomena and the simulation of a wider range of problems. For instance, incorporating features like single-qubit gates or enabling single-site addressability enhances the level of control over the quantum simulation. These additional capabilities not only improve the precision of simulations but also extend the applicability of analog quantum simulators to solve a more extensive set of quantum information processing tasks. Analog quantum simulators find application across various quantum platforms, including superconducting qubits\cite{biblatexsubmittingtothearxiv}, trapped ions \cite{arxivpdfoutput}, and ultracold atoms \cite{roch}, where they effectively mimic the behavior of physical quantum systems. This diversity in platforms underscores the versatility and potential impact of analog quantum simulation in advancing our knowledge about quantum phenomena and facilitating the applications of quantum information technologies.

On the other hand, digital quantum simulators \cite{nourda} allow for analog-like elements, such as switching multiqubit interaction on and off, rather than decomposing multi-qubit gates into single and two-qubit gates \cite{nourda}. This study focuses on digital quantum simulation using analog-digital  techniques to provide a more realistic path to solving scientifically important and hard problems in the near term \cite{proof}. To accurately mimic and understand a quantum system described by the Hamiltonian, we aim for a precise approximation and its implementation in quantum superconducting circuits. As we explore the model, we pay close attention to the dynamics, especially when $g^R$ gets comparable to the energy assocaited with the bosonic mode $\omega_b^R$.

In our pursuit of achieving a high-fidelity realization and emulation of the Hamiltonian, we strive for a more accurate approximation and the implementation of a targeted Hamiltonian in quantum superconducting circuits. Navigating through the complexities of the model, we unravel intricate dynamics, particularly when the coupling strength $g^R$ is larger than the energy of the bosonic mode $\omega_b^R$.

Our approach critically evaluates the feasibility and efficiency of digital simulation to comprehend the system's behavior, especially as it evolves in deep strong coupling intensities in the contexts of superconducting circuits. In these scenarios, the coupling strength becomes reaches or even becomes higher than the frequency of a specific bosonic mode. We delve into the challenges associated with the coherence of measurements over extended periods in the realm of superconducting circuits. Moreover, we explore the advantages of employing digital simulation techniques to extract the system's behavior, a task that has proven challenging to achieve experimentally for many years.

By leveraging the capabilities of digital simulation, we aim to gain a deeper understanding of the system's intricate dynamics during prolonged periods in superconducting circuits, where traditional experimental approaches face limitations. Our investigation not only sheds light on the challenges posed by strong coupling regimes but also highlights the benefits of digital simulations in elucidating behaviors that have remained elusive through experimental means for an extended period.

We organise our article as: In Sec. \ref{sec2}, we give detailed trotterization method, while the implication of digital simulation is given in Sec. \ref{sec3}. The results and discussed in Sec. \ref{sec4}, while summarized in Sec. \ref{sec5}. Finally, we give the version, tables and appendix related to our study in Secs. \ref{sec6}, and \ref{sec7}.

\section{Trotterization}\label{sec2}
The digital quantum simulation relies on Trotterization \cite{arranded} method. A quantum simulator encounters provide solutions to the challenges in direct replication of the dynamics, digital quantum simulations regarding a quantum system. These quantum simulations offer a solution by enabling the engineering of arbitrary quantum dynamics through the decomposition of the original evolution into implementable gates, a process, namely, the Lie-Trotter-Suzuki decomposition. The named approach undertakes the time unitary operator considering a specific Hamiltonian while incorporating some implementable quantum gates.
{

Besides, the Trotter expansion is approximated through a specific formulation, with the number of Trotter steps playing a crucial role in achieving accurate digitized dynamics. Simulating longer times requires a larger number of digital steps to ensure fidelity.

next, for the time evolution while assuming an initial state \(|\psi(0)\rangle\) can be can be done by the Schr\"{o}dinger equation:
\begin{equation}
|\psi(t)\rangle = U(t) |\psi(0)\rangle,
\end{equation}

where \(U(t) = \exp{(-iHt})\)  represents the unitary time evolution operator associated with the system's Hamiltonian \(H\). The Trotter expansion of the unitary operator \(e^{-iHt}\) can be approximated as:
\begin{equation}
e^{-iHt} \approx \left( e^{-iH_1 t/l} \cdot \ldots \cdot e^{-iH_M t/l} \right)^l + \sum_{i<j} \frac{[H_i, H_j]t^2}{2l}.
\end{equation}

In this formulation, the number of Trotter steps in total are represented by \(l\). Note that increasing \(l\) enhances the accuracy of digital simulations. As indicated in Eq. (2), the primary error part scaling with \(t^2/l\) and is influenced by the commutators of the involved Hamiltonians, which are proportional to the relatice coupling constants, and norms. Consequently, simulating prolonged` times necessitates a larger number of digital steps to ensure satisfactory fidelities.

The generalized Trotter's formula \cite{arranded} can be symmetrized \cite{sym} to take the form:
\begin{widetext}
    \begin{equation}\label{eq:generalized_trotter}
    e^{\hat{H}_1 + \hat{H}_2 + \cdots + \hat{H}_p} = \lim_{l\rightarrow \infty} \left(e^{\hat{H}_1/2l}e^{\hat{H}_2/2l}\cdots e^{\hat{H}_{p-1}/2l}e^{\hat{H}_p/l}e^{\hat{H}_{p-1}/2l}\cdots e^{\hat{H}_2/2l} e^{\hat{H}_1/2l}\right)^l.
\end{equation}
\end{widetext}

This symmetrized Trotter's formula exhibits a correction of \(l^{-2}\). The advantage of employing this symmetrized version lies in its faster convergence compared to the unsymmetrized formula, while maintaining a similar precision for pulsed Hamiltonians.

\section{The digital quantum simulation model}\label{sec3}

In the Quantum Rabi model, in particular regarding the notion of cQED, we assume a 2D atom initially prepaed in a transmon qubit state, influenced by a microwave resonator \cite{rabi, tran}. The relative Hamiltonian (with $\hbar = 1$) for the discripted model can be written as:
\begin{equation}
    \hat{H} = \frac{{\omega_Q}}{2} \sigma^z + \omega_b^R \hat{b}^\dagger \hat{b} + g^R \sigma_x (\hat{b}^\dagger + \hat{b}),
\end{equation}

where $\omega_Q$ represents the qubit energy splitting,  $\omega_b^R$, and $g^R$  denote the bosonic mode frequency and transversal coupling strength. Here, $\sigma^i$ denotes Pauli matrices with $\sigma^x = (\hat{\sigma}_+ + \hat{\sigma}_-)$, and $\hat{\sigma}_z = |e\rangle\langle e| - |g\rangle\langle g|$ represents the eigenstates of the relative basis.

Additionally, $\hat{b}^\dagger$($\hat{b}$) are creation(annihilation) operators acting on the Fock space, respectively. We consider a small geometric coupling $g^R \ll \{\omega_Q,~ \omega_b^R$\}, such that to allow deploying the Rotating Wave Approximation (RWA), resulting in the Jaynes–Cummings Hamiltonian form \cite{rabi}:
\begin{equation}
    H = \frac{\omega_Q}{2}\sigma_z + \omega_b^R b^\dagger b +  g^R( b \sigma^\dagger  +b^\dagger  \sigma ).\label{eq:hamiltonian}
\end{equation}
Note that the Hamiltonian has no counter-rotating terms, however, with the condition to preserve the excitations numbers.

Next, to implement a hybrid digital-analog model based on the full Quantum Rabi model, encoding the modes of the atom as well as of the field, in a transmon quantum system and coplanar-waveguide resonator case, one can utilize local rotations $\sigma_x H_J \sigma_x$. This allows for encoding two similar Jaynes-Cummings interactions through digitalization, represented as $H_R = H_{JC} + H_{AJC}$, associated with this model. The $H_{JC}$ , and $H_{AJC}$ are given by:
\begin{align}
    H_{JC} &= \frac{\omega_b^R}{2} b^\dagger b + \frac{\omega_1^Q}{2} \sigma_z + g(b^\dagger \sigma^- + b \sigma^+), \\
    H_{AJC} &= \frac{\omega_b^R}{2} b^\dagger b - \frac{\omega_2^Q}{2} \sigma_z + g(b^\dagger \sigma^+ + b \sigma^-),
\end{align}

In the realm of superconducting circuits, we propose a bivariate modulation approach involving two simultaneous steps to control the rate of qubit transition frequency. Additionally, we introduce the $\tilde{\omega}_{RE}$ frame, a pivotal constructive step in achieving deep strong coupling with anharmonic transmon qubits. This rotating frame assumes significance by affording control over simulated frequencies, specifically $\omega_Q$, and $\omega_b^R$. Conventionally, the rotating frame frequency  is ascertained using a practical generator or some chosen device signal, serving to define a rotation or measurement basis.

In the Trotterization realizations, the rotating frame conserves the relative abstract nature, lacking a specific drive for effective frequency control. To remedy this, we advocate the employment of a high-resolution and flexible effective Hamiltonian, keeping the abstract rotating frame with precision and versatility. Consequently, this approach holds promise for the nuanced manipulation of quantum states within the domain of superconducting circuits.

The effective interaction Hamiltonian in this scenario is expressed as follows:
\begin{equation}
    \tilde{H_{eff}} = \tilde{\Delta}_b b^\dagger b - \tilde{\Delta}_Q \sigma_z/2 + g(b^\dagger \sigma^- + b \sigma^+),
\end{equation}
where $\tilde{\Delta}_b = (\omega_b - \tilde{\omega}_{RE})$ and $\tilde{\Delta}_Q = (\omega_Q - \tilde{\omega}_{RE})$. Therefore, Eq. (4) can be assumed equivalent to $H_{JC}$ while redefining the associated coefficients. The  terms \(H_{AJC}\) causing counter rotation is simulated while deploying a qubit rotation to \(H_{JC}\) and setting a unique detuning strength of the transition frequency of the qubit as:
\begin{equation}
    \sigma_xH_{JC}\sigma_x = \tilde{\Delta}_rb^\dagger b - \tilde{\Delta}_Q\sigma_z + g(b^\dagger\sigma^+ + b\sigma^-). \label{eq:counter_rotating}
\end{equation}

By strategically tuning the qubit-resonator detunings in two distinct steps: \(\tilde{\Delta}_{iQ}\) (with $i\in\{1,2\}$) represent the initial phase, and subsequent final rotation. In this regard, we analyze the capability of the considered configuration to simulate the Hamiltonian of a Quantum Rabi system, as outlined in Eq. (5). This approach of digitization, as shown in Eqs. (2), and (3), involves skillful exploitation of the simulated interactions.  Thus, showcasing a sequence where standard resonant Jaynes-Cummings interactions, characterized by varying qubit transition frequencies, are characterized and regulated by precisely timed microwave pulses  \cite{biblatexsubmittingtothearxiv}. These pulses serve the purpose of executing routine qubit flips \cite{example}. Employing this sequence in regard to the considered digitization scheme gives optimal characteristics and solution of the intricate dynamical map of the Quantum Rabi Model.

Through the standard Trotterization applied to the Quantum Rabi Model, we derive effective parameters that illuminate the underlying physics. The coupling between the resonator and the bosonic frequency is elegantly expressed as \(g_R = g\), with \(g\) regulating the strength of the coupling while the detuning of the resonator frequency is defined as \(\omega_{Rr} = 2\tilde{\Delta}_r\). Therefore, making a connection between the frequency of the transitions of two levels with two sequential steps, given by \(\omega_{RQ} = \tilde{\Delta}_{1Q} - \tilde{\Delta}_{2Q}\).

In the pursuit of achieving high fidelity in digital quantum simulations, we set \(\tilde{\Delta}_{2Q} = 0\). This choice, we believe will minimize the loss in simulation accuracy, ensuring that our digital simulation remains robust and reliable, paving the way for groundbreaking insights into the dynamics of the Quantum Rabi Model.

\section{Fidelity simulations}\label{sec4}
Here, we explore fidelity dynamics in the Quantum Rabi Model across a range of considerations, analyzing the fidelity when characterized by various inclusive parameters. The visual representations shows valuable insights into the accuracy and convergence of our approximations done on Quantum Rabi Model. One can define the fidelity as:
\begin{align}
     &\max\{0, 1 - \|\hat{\mathcal{U}}_m(T)-\hat{U}(T)\|\} \le \nonumber\\
     &\left|\langle\psi|\hat{U}^\dagger(T)\hat{\mathcal{U}}_m(T)|\psi\rangle\right| \le 1.
    \end{align}

Using the above relation, we consider fidelity as the overlap between the unitary exact Hamiltonian and the unitary approximated Hamiltonian under the general and symmetrized Trotter formula in Eqs (2), and (3). in particular, we focus on a lower bound that does not depend on the initial given states. 
\section{Configuration and parameter settings}
Next, utilizing fidelity, we explore the behavioral dynamics by first assuming an atom as transmon qubit with a high-frequency transition which is excited as: $\omega_1^Q=2\pi \times 6.381$ GHz and $\omega_2^Q=2\pi \times 5.452$ GHz \cite{biblatexsubmittingtothearxiv} while in the next, we assume $g=2\pi \times 1.79$ GHz. For the resonator frequency, we explore different values to understand the system's behavioral dynamics.

In our study, we investigated the impact of varying the maximum photon population of the Bosonic mode in the simulations, as indicated by the measured dispersive shift of the readout resonator. We considered a photon population of $100$ in Figs. \ref{fig:a}-\ref{fig:f} and in Figs. \ref{fig:symmetrized11}-\ref{fig:symmetrized55}, taking $30$ photon populations for different simulation durations denoted as $T = 30\text{s}$ and $T = 2\text{s}$. The following analysis is aimed to investigate the temporal effects in measurements and the coherence time between the considered quantum states \cite{z, x, v}. Additionally, we intend to understand the ratio of system-resonator coupling $g/\omega_b^R$.

Before going into the detailed description of the figures, note that the simulations were initiated with the resonator in the initial state $|0\rangle$, and the qubit in states $|e\rangle$, $|g\rangle$, $|+\rangle$, and $|-\rangle$. Our findings demonstrated a correlation between the resonator and qubit, consistent with the anticipated Bell-cat entanglement (entanglement between qubit and photon in superconducting qubits). Specifically, the simulation results for the expected Bell-cat state revealed an outcome of $|e(g)\rangle$ for the qubit, leading to the resonator being in an odd (or even) Schrödinger cat state $ |+,+\alpha\rangle_{Q,b} -  |-,-\alpha\rangle_{Q,b}$. Besides, we also simulated a higher number of steps $l$,  which interestingly, resulted into enhanced accuracy in the simulation outcomes.

\section{Main Results and Discussion}
In this section, we utilize fidelity as a function of the number of steps $l$, as shown from Figs. \ref{fig:symmetrized11}-\ref{fig:symmetrized55}. We find an overlap between the initial state evolved under the exact unitary Hamiltonian and the unitary approximated Hamiltonian using the general and symmetrized version Trotter formula, where the initial states are assumed to be \(|i\rangle = |0, e\rangle\).
\begin{figure}[ht]
    \centering
    \includegraphics[width=1\linewidth]{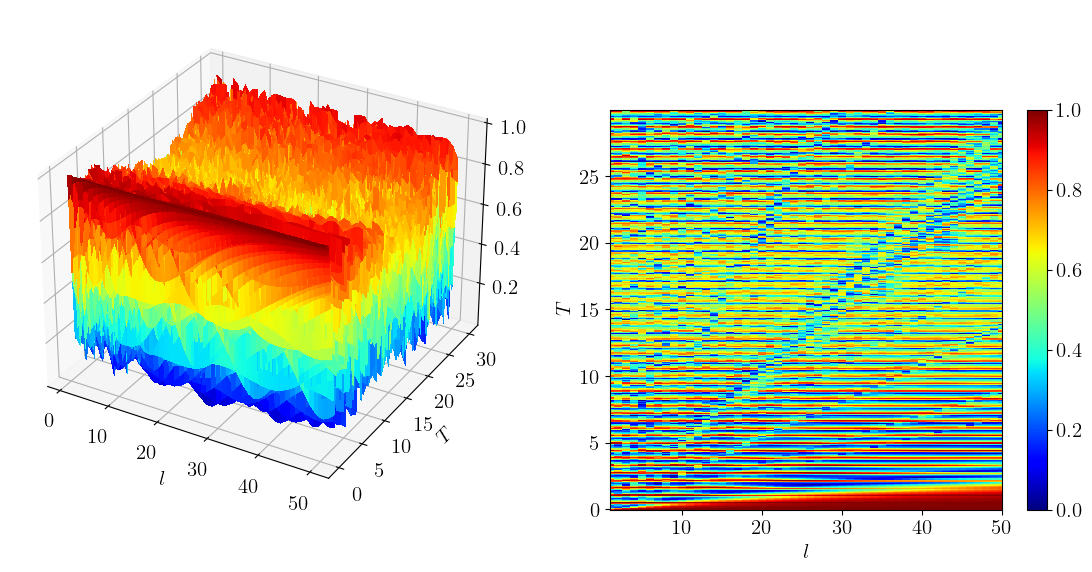} 
    \includegraphics[width=1\linewidth]{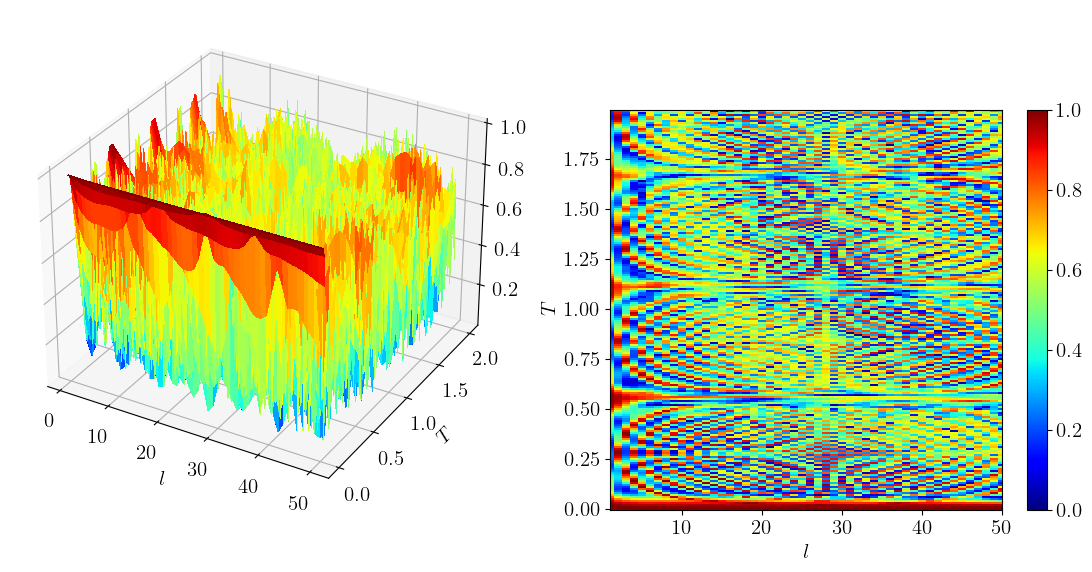}
    \caption{Upper panel: The plot illustrates the fidelity as a function of the number of steps $l$ for various values of $\omega T/2\pi$. The inset offers insight into the overlap between the ideally evolved state and the initial state, where all qubits are in the state $|i\rangle \equiv |0, e\rangle$. This depiction utilizes the general Trotter formula to illustrate the overlap. For $T = 30 s$, Eq. (2) showcase both the exponentially exact unitary Hamiltonian and the approximate unitary Hamiltonian under the general Trotter formula. In the first order, these equations are applied to $H_{JC}$ and $H_{ANJ}$ with $g = \omega_b^R$, involving 30 photons in the resonator. Lower Panel: Same as upper panel but for $T = 2s$ .}
    \label{fig:symmetrized11}
\end{figure}
In Fig. \ref{fig:symmetrized11}, the upper and lower panel initially shows that the fidelity of the system remain maximum. However, with the increase in time $T$, the fidelity shows drop from the maximum level.

Besides, one can see an evident oscillatory behavior, meaning the exchange of information between the qubit and resonator. Most importantly, for $T=30 s$ in the upper panel, the fidelity slopes touces the maximum bound at regular intervals. However, for $T=2 s$, mostly the fidelity slopes reaches the maximum bound either at lower ($l<10$) or higher steps ($l>40$).  It seems that fidelity corresponds to coarseness in time. However, in the left Fig. \ref{fig:symmetrized11}, they illustrate Schrödinger cats over time evolution.

\begin{figure}[ht]
    \includegraphics[width=1\linewidth]{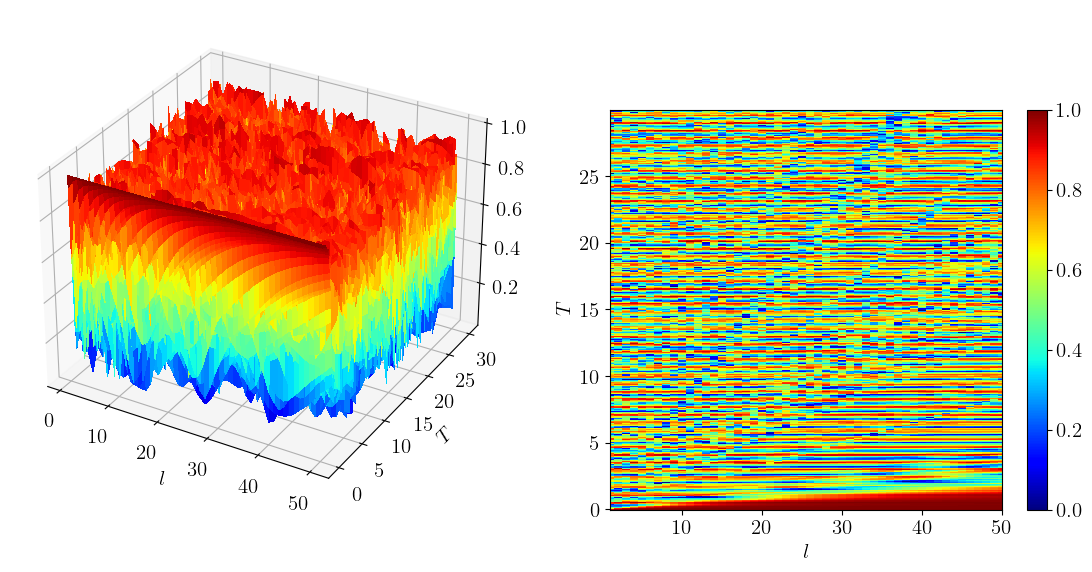} \put(-200,0){}]
    \includegraphics[width=1\linewidth]{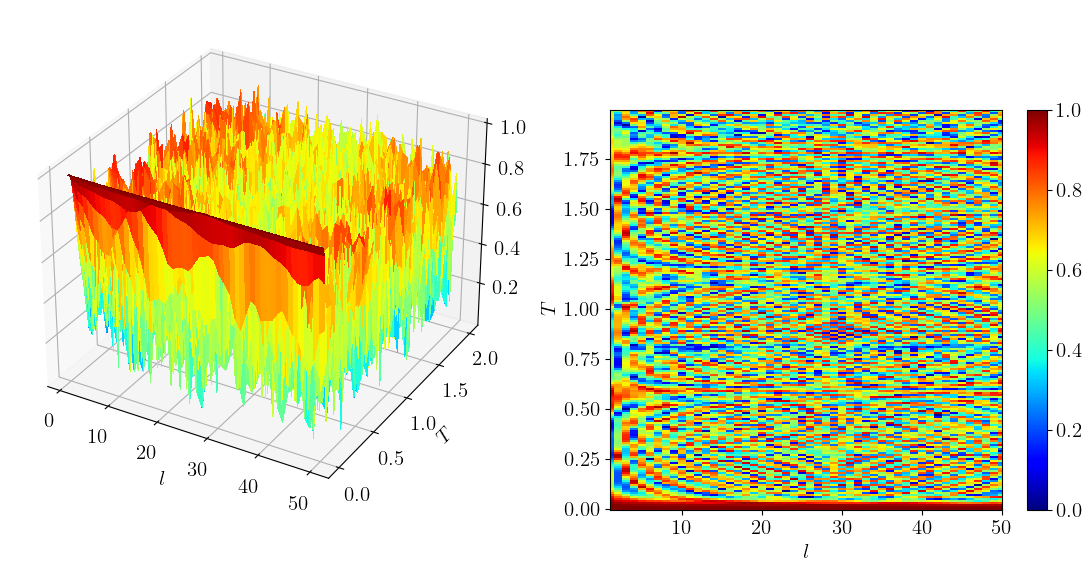}\put(-200,0){}
    \caption{upper panel: The plot illustrates the fidelity as a function of the number of steps $l$ for various values of $\omega T/2\pi$. The inset provides insight into the overlap between the ideally evolved state and the initial state, where all qubits are in the state $|i\rangle \equiv |0, e\rangle$. This visualization of overlap utilizes the general Trotter's Formula. For $T = 30 s$ , Equation 2 showcases both the exponentially exact unitary Hamiltonian and the approximate unitary Hamiltonian under the general Trotter formula. In the first order, these equations are applied to $H_{AJC}$ and $H_{JC}$ with $g=\omega_b^R$, involving 30 photons in the resonator. Lower Panel: Same as upper panel but for $T = 2$ns.}
    \label{fig:symmetrized33}
\end{figure}

Besides, compared to Fig. \ref{fig:symmetrized11}, a good agreement can be seen  with Fig. \ref{fig:symmetrized33} while showing similar results, showcasing deep strong coupling in superconducting circuits corresponding to Schrödinger cats. Instead of exponentially rising with the order \({H_{JC}, H_{ANJ}}\), we changed the order to \({H_{ANJ}, H_{JC}}\). The results are consistent in Figures \ref{fig:symmetrized33} as in Fig. \ref{fig:symmetrized11}. After analyzing this result, we can conclude that fidelity exhibits a long coarseness in time, supporting higher gate fidelity and showing the capacity to serve as quantum memories in superconducting circuits.

\begin{figure*}[ht]
    \centering
    \includegraphics[width=0.7\linewidth]{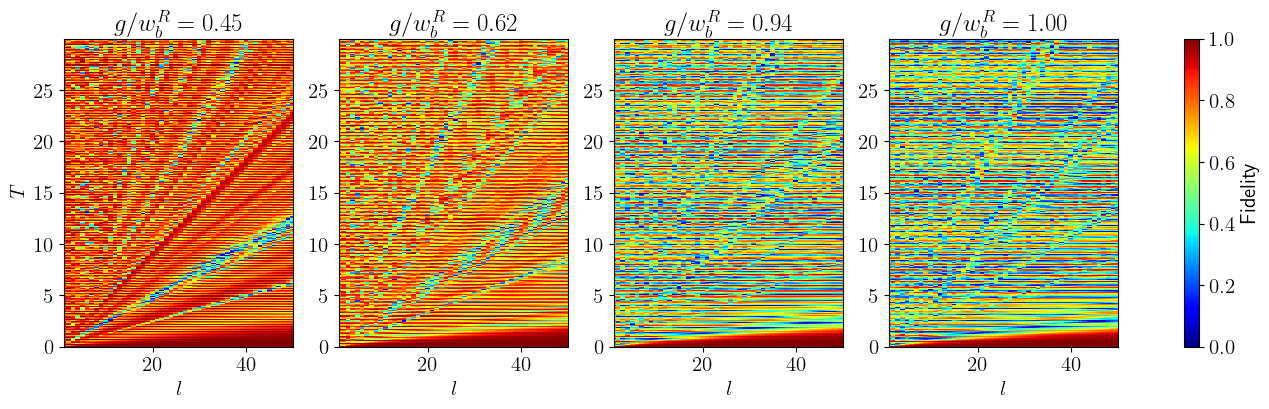} \quad
     \includegraphics[width=0.7\linewidth]{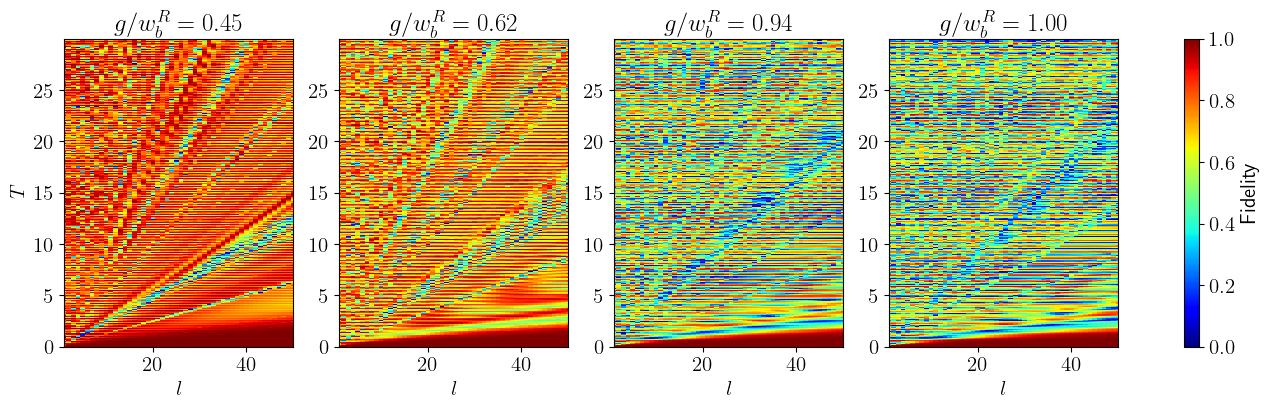}
    \caption{Upper panel: The plot depicts the fidelity as a function of the number of steps $l$ for different values of $\omega T/2\pi$. The inset provides a visual representation of the overlap between the ideally evolved state and the initial state, where all qubits are in $|i\rangle \equiv |0, e\rangle$. This representation of overlap of initial states  utilizes the general Trotter's Formula with $30$ photons in the resonator and varying bosonic mode energy for different coupling ratios $g/\omega_b^R$. Lower Panel: Same as upper panel but for symmetrized Trotter's Formula.}
    \label{fig:symmetrized55}
\end{figure*}

After analyzing the upper panel of Fig. \ref{fig:symmetrized55}, we observed that when the coupling ratio $g/\omega_b^R$ is smaller, the digital error decreases. However, when the coupling strength $g$ is much larger than the bosonic mode energy $\omega_b^R$, the fidelity starts to exhibit a digitally increasing error. This is evident in the figure as it starts from the origin and originates from the commutator between $H_{ANJ}$ and $H_{JC}$, and the quantum logic gate. By optimizing control pulse sequences to suppress the impact of environmental noise, transmon qubits can achieve significantly higher quantum gate fidelities, paving the way for more robust quantum computations. Furthermore, the lower panel of Fig. \ref{fig:symmetrized55} shows a similar behavior, but the results converge more quickly compared to the upper panel.

Next, we present a comprehensive analysis of dynamic solutions and digital quantum simulations applied to the quantum Rabi model within superconducting circuits, as depicted in Fig. \ref{fig:a}-\ref{fig:f}. Each figure intricately captures the significant factors about the quantum system under investigation. The blue line within each figure analyze the qubit population evolution of the driven system, showcasing both the exact and digital quantum simulations. Concurrently, the green line illustrates the bosonic population of the quantum Rabi model, denoted by $\langle b^\dagger b \rangle(t)$, as it evolves over time $\Delta t$.

\begin{figure*}[ht]
    \centering
    \includegraphics[width=0.45\linewidth]{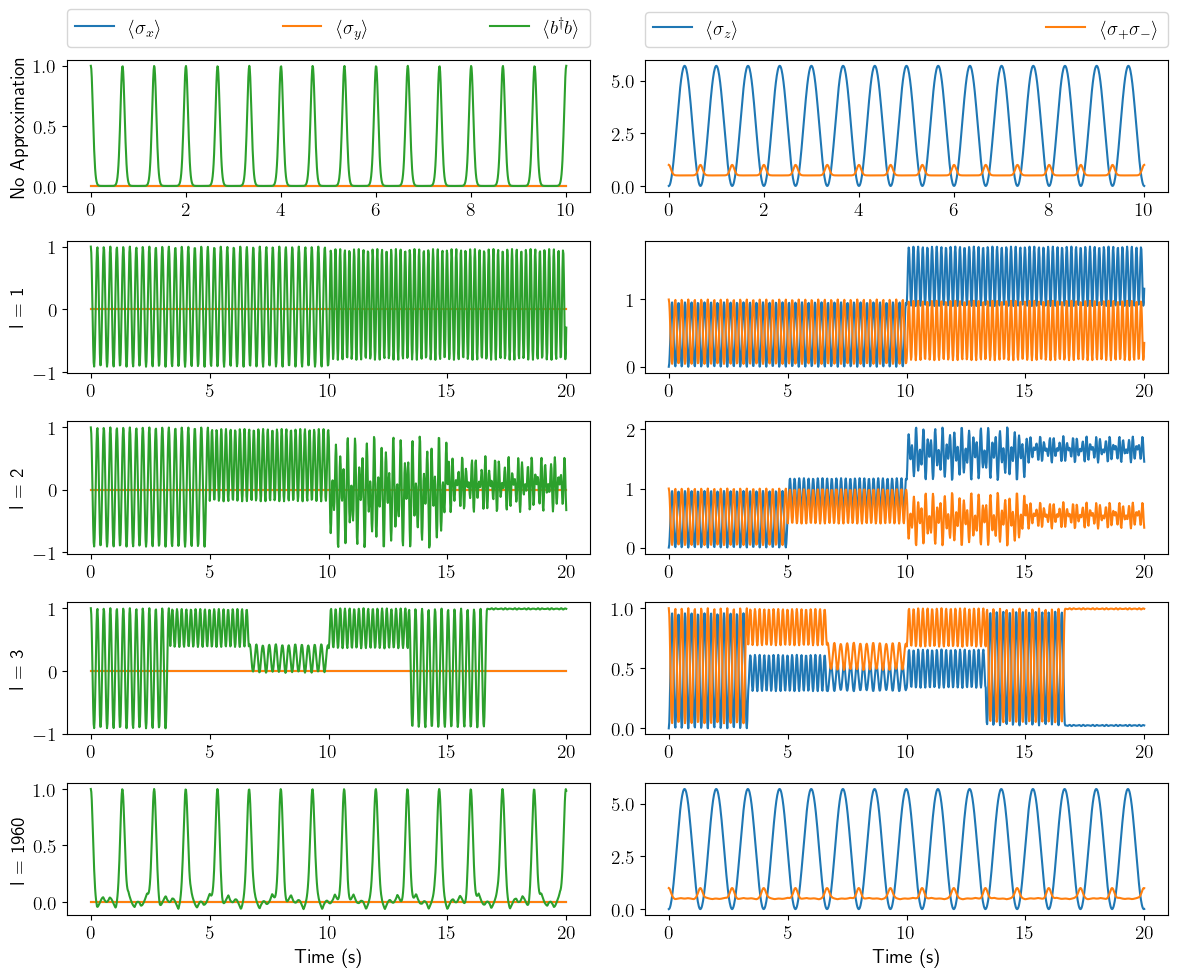} \quad
    \includegraphics[width=0.45\linewidth]{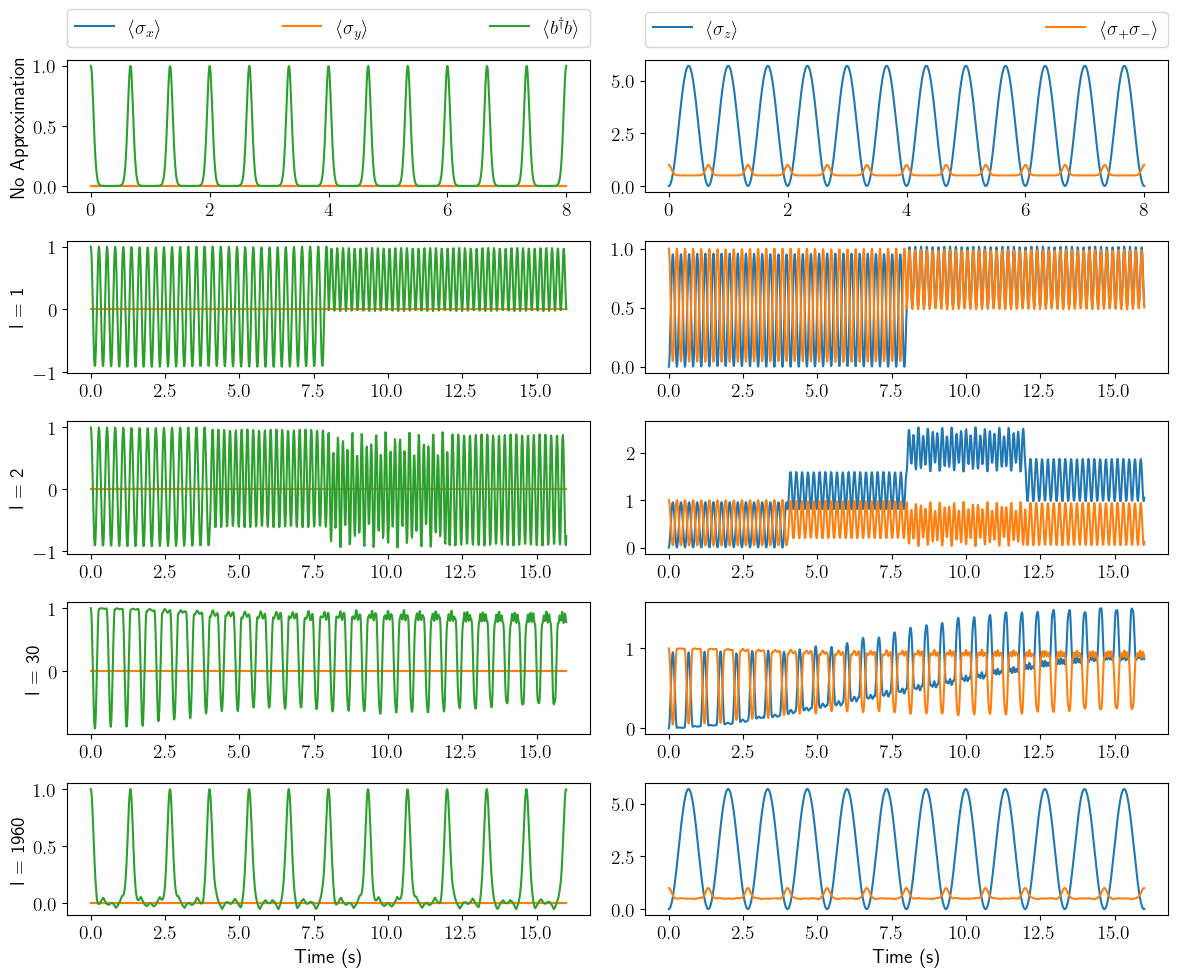}
    \caption{Left side: Quantum Simulation of the Quantum Rabi Model. The figure illustrates the comparison between exact solutions and digital simulations using the general Trotter formula. The simulations were performed with different numbers of steps (\(l\)) in the degenerate case were ( $\omega_Q^R=0$) with $g=w_b^R$. Right Side: Same as left side but for symmetrized Trotter formula.}
    \label{fig:a}
\end{figure*}

\begin{figure*}[ht]
    \centering
    \includegraphics[width=0.45\linewidth]{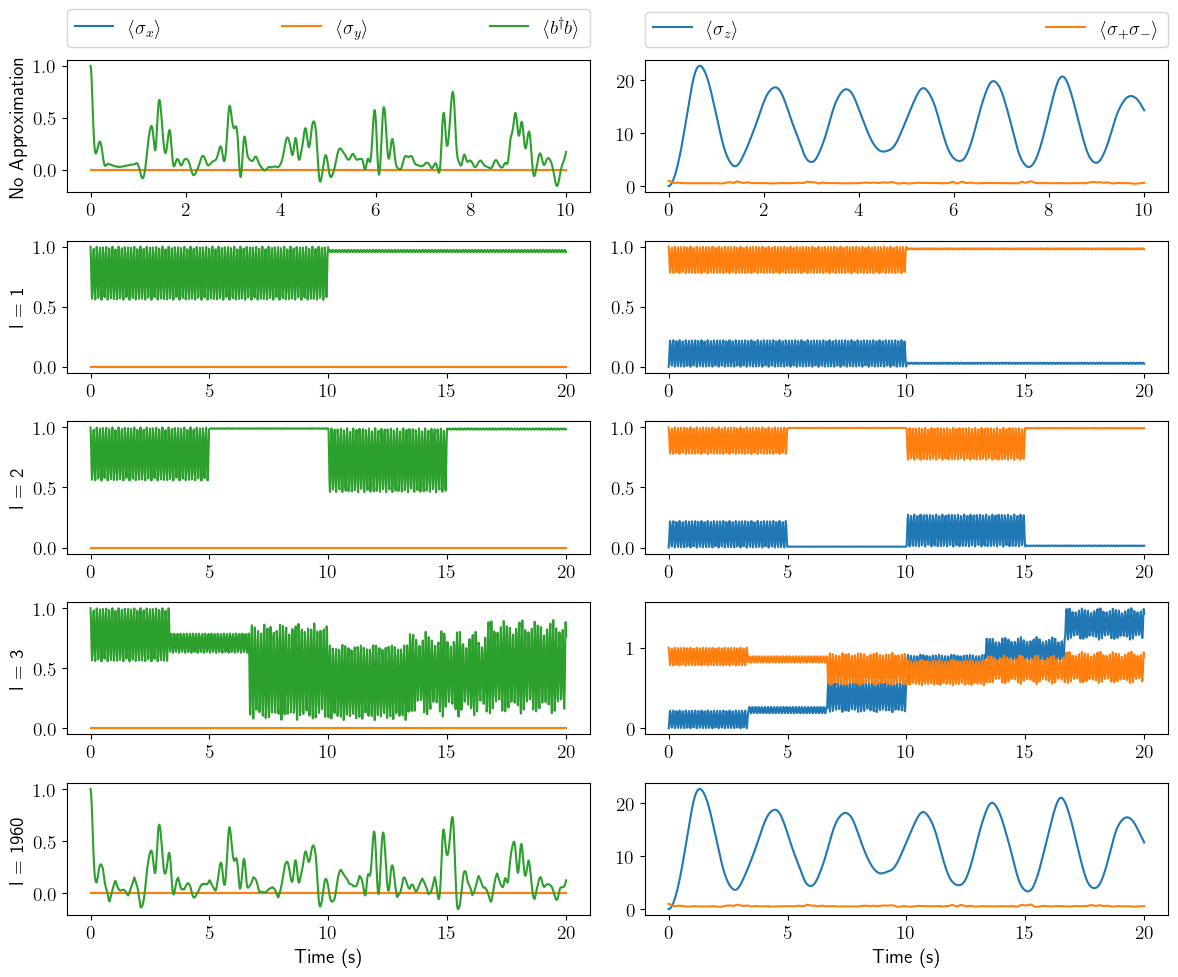} \quad
     \includegraphics[width=0.45\linewidth]{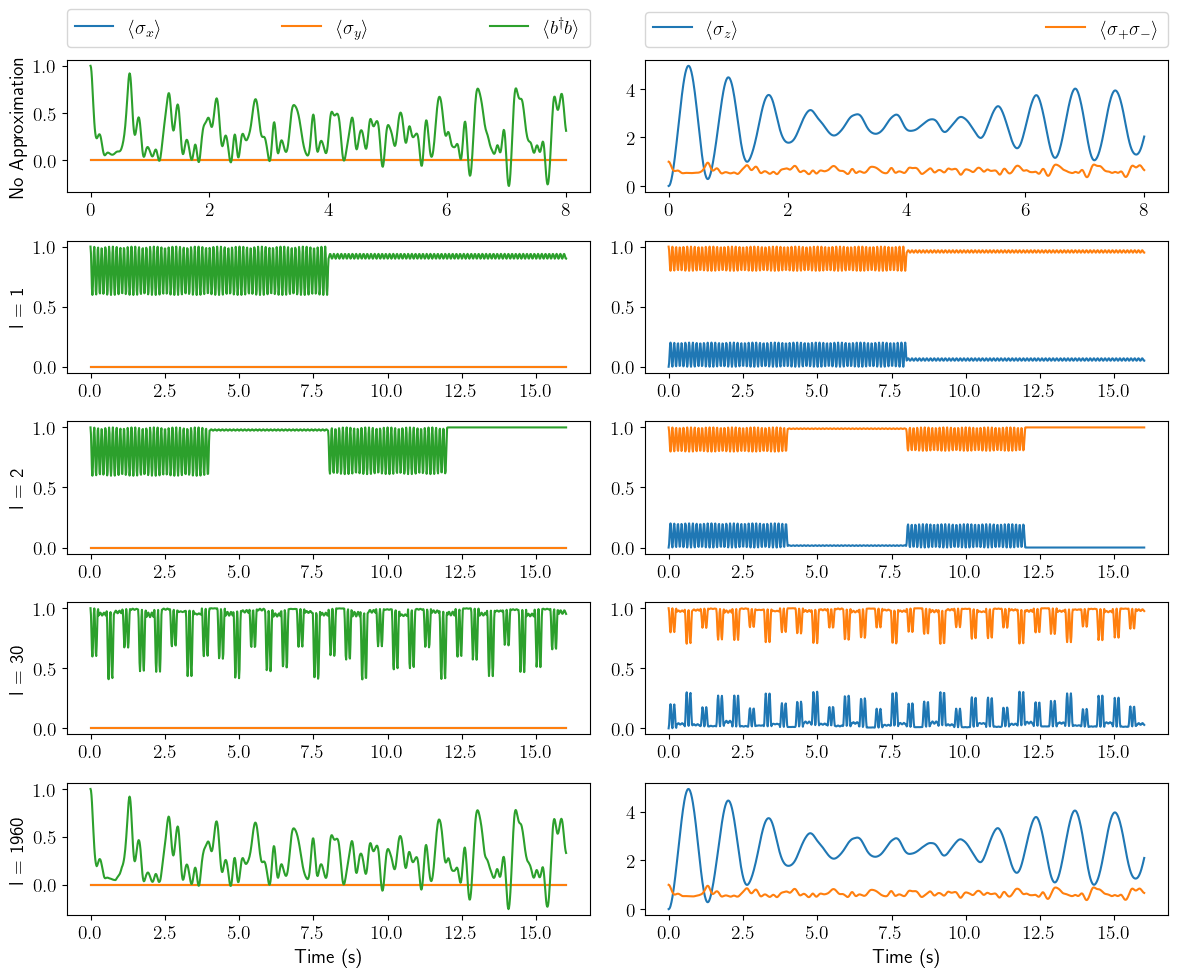}
    \caption{Left side: Quantum Simulation of the Quantum Rabi Model. The figure illustrates the comparison between exact solutions and digital simulations using the general Trotter formula. The simulations were performed with different numbers of steps (\(l\)) under the condition $g/\omega_b^R\approx 2.67$.  Right Side: Same as left side but for symmetrized Trotter formula with coupling ratios $g/\omega_b^R\approx 1.193$ }
    \label{fig:c}
\end{figure*}

\begin{figure}[ht]
    \centering
    \includegraphics[width=1\linewidth]{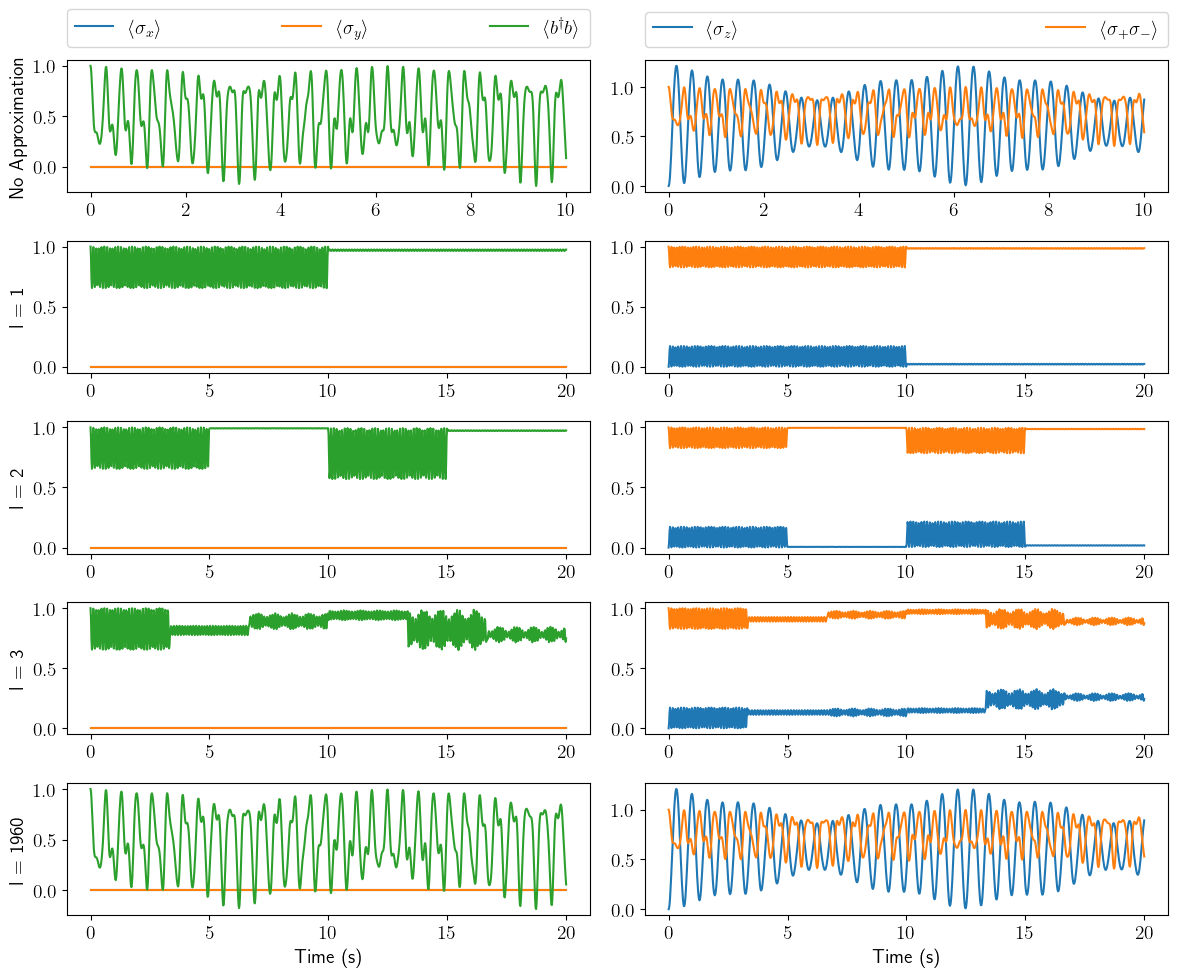}
    \caption{Quantum Simulation of the Quantum Rabi Model. The figure illustrates the comparison between exact solutions and digital simulations using the general Trotter formula. The simulations were performed with different numbers of steps (\(l\)) under the condition \( g/\omega_b^R \approx 0.62 \).}
    \label{fig:f}
\end{figure}

Notably, the orange line presents the spin average measurement, $\langle \sigma^\dagger \sigma_- \rangle(t)$ providing valuable insights into the quantum dynamics under investigation. The simulations conducted here include a deep exploration of various initial states. Our methodology aims to reach an understanding of population evolution in terms of deep strong coupling. Besides, our results shows a restricted case with zero atomic frequency $(\omega_Q^R=0)$ to demonstrate key existence and verification of the simulation of deep strong coupling. These signatures include characteristic collapses and revivals in both atom and resonator parities. Moreover, the coherent oscillations in resonator population reaching large photon numbers, as well as the opposition between the cavity phase-space trajectories are also evident.

Specifically, in Figs. \ref{fig:c}, and \ref{fig:f}, we examine the qubit time evolution for scenarios where the system is characterized by $\omega_b^R/(2\pi) \approx 0.67$, $1.5$, and $2.9$ GHz. Correspondingly, the dimensionless coupling strength $g/\omega_b^R$ assumes values of approximately $2.67$, $1.19$, and $0.62$, respectively. In the above plots, we conducted a comprehensive exploration of quantum Rabi dynamics ranging from the ultrastrong to extremely deep strong coupling regimes. When the Jaynes-Cummings interaction is assumed in the ultrastrong regime, terms such as $b \sigma$ and $b^\dagger \sigma^\dagger$ cannot be neglected, causing the total parity to be conserved. Without the strong symmetry of number conservation, solving the full Quantum Rabi Model (QRM) becomes challenging. 
However, predicting these phenomena accurately is particularly challenging because they involve complex quantum interactions and dynamics. Quantum systems exhibit rich behavior and can undergo intricate transformations, making it difficult to analytically determine the properties of their ground states. Additionally, experimental observations of these phenomena are still limited, adding to the complexity of predicting their exact characteristics.

Using a master equation as indicated under exact and approximate unitary Hamiltonian scenarios, we analyze in Fig. \ref{fig:a} the degenerate case \(\omega_Q^R = 0\) (where the related solution is given in Appendix B). Note that the degenerate case has been thoroughly investigated for the production of Schrödinger cat states, as evidenced in Fig. \ref{fig:a}. Coherent oscillations in the resonator population are observed when varying the maximum photon, and similarly for the average \(\langle b^\dagger b \rangle\). These coherent oscillations are shown under exact simulation and digital quantum simulation with a higher number of steps. The results demonstrate a similarity, indicating good agreement and verifying a deep, strong signature. Furthermore, simulated deep strong coupling demonstrated in our configuration leads to conditionally non-classical Schrödinger cat states in the resonator. This verifies the presence of qubit-resonator entanglement arising from coherent deep strong coupling dynamics.

In the non-degenerate case, where $\omega_Q^R \neq 0$, the bosonic modes exhibit a specific population behavior. The population of these modes reaches a maximum and then reverts to its initial state when the frequency is $\omega_b^R / (2\pi)$ at some given interval points, even under the absence of dissipation.

Now, we compare the exact solution to those obtained through digital quantum simulation in Figs. Figs. \ref{fig:a}-\ref{fig:f}, using both symmetrized and general Trotter formulae. It is observed that the exact solution and digital simulation  both indicate the population of bosonic modes $\langle b^\dagger b \rangle$ reaching its maximum faster when the energy of the bosonic mode, denoted as $w_b^R$, is larger. This corresponds to a smaller coupling ratio $g/\omega_b^R$. Consequently, a digital quantum simulation with an increasing number of steps demonstrates a behavior that aligns closely with the exact simulation. Most importantly, due to practical flux-pulsing bandwidths which limit the shortest achievable Trotter step, makes it challenging to digitize fast, compared with the dynamic simulation. Furthermore, reaching acceptably low Trotter error in the interesting regimes of $g/\omega_b^R$ requires small qubit-resonator coupling. This also places constraints on other configuration parameters, including the long simulation times as seen in Figs. \ref{fig:a}-\ref{fig:f}.

In Fig. \ref{fig:c}, we present both an exact solution and a digital quantum simulation depicting the dynamics of qubit population. Interestingly, a noticeable similarity emerges as the number of simulation steps increases. Particularly during Jaynes-Cummings interactions, there is a coherent influx of photons into the resonators. However, in the simulation, qubit flips significantly amplify photon production and increase coherence over a specified time period.

In Fig. \ref{fig:f} (left side), a decay in the prediction of photon count is evident over a certain short duration. This decay takes some time, and subsequently, an enhancement in photon prediction is observed under qubit flip. Interestingly, in Fig. \ref{fig:f} (right side), this enhancement is much faster, and the prediction exhibits coherence over time. Moreover when  a bosonic mode $\omega_b^R$. Besides this, it seems larger when varying the maximum photon count in the resonator conduct an analysis of the influence of increasing $\omega_b^R$ on augmenting the energy of the bosonic mode.  Moreover, the resultant digital quantum simulations show similar behavior when number of steps is set larger.

 \section{Conclusion}\label{sec5}
In this article, we present a comprehensive exploration of the critical aspects of validating the behavior of Quantum Rabi model and associated complex digital simulations poised to achieve quantum advantage~\cite{examplecitation}. By employing both general and symmetrized Trotter formulas with an increased number of steps, we illustrate a striking resemblance between digital simulators and exact dynamic simulations in superconducting circuits. Our approach involves plotting fidelity under various conditions, choosing smaller Trotter steps, modifying time evolution and coupling strength to enhance simulator performance.

Importantly, our findings reveal a departure from the limitations imposed by the error-per-gate case seen in previous cQED simulations. This departure empowers us to linearly increase the number of steps for extending simulated time, thereby, showing an advantage of quantum simulators in sustaining long coherent times. This newfound flexibility not only augments the simulator's capacity but also allows for a more accurate approximation of the full quantum Rabi model.

Moreover, the results indicate that the simulator's efficacy is not bound by any fixed error threshold per gate, as observed in earlier cQED simulations. This breakthrough enables us to progressively scale the number of steps, resulting in a quantum advantage characterized by prolonged coherent times. The improved performance and enhanced flexibility presented in our study mark a significant stride toward leveraging the full potential of quantum simulators in complex digital environments. Even a single decay destroys qubit-resonator entanglement, and losing a photon becomes increasingly likely for larger photon numbers. Thus, this article provides proof through exact and digital quantum simulation that qubit flips significantly amplify photon production and increase coherence over a time evolution that supports an exploration of entanglement in qubit-resonator coupling, where the coupling strength $g$ dominates over subsystem energy.

\bibliographystyle{plain}

\subsection{Appendix B}
Inaccuracies of the implementation of the model on a quantum computer can stem from different causes of completely different nature. We will first consider errors that arise from the Trotterization of the evolution . We will then consider errors due to the noisy nature of the quantum computer ,The dissipative part of the dynamics is here described by a Markovian master equation in Gorini-Kossakovski-Sudarshan-Lindblad form \cite{g}:

\begin{equation}
\frac{d\hat{\rho}}{dt} = -\frac{i}{\hbar} [\hat{H}, \hat{\rho}] + \gamma \sum_k \left(2\hat{L}_k\hat{\rho}\hat{L}_k^\dagger - \{ \hat{L}_k^\dagger \hat{L}_k, \hat{\rho} \}\right)
\end{equation}

Additionally, we consider Reduced-Noise Models to explore the potential performance of future hardware with lower noise levels. For this purpose, we employ the same error channels that IBM uses to characterize their current quantum devices.

For the specific simulations, we set $\gamma = 1$. With these parameters, the accurate evolution of the system up to a specific time is indicated in the figures.
\end{document}